%% file: main.tex
\documentclass{sigchi-ext}
\usepackage[T1]{fontenc}
\usepackage{textcomp}
\usepackage[hypcap = false, hypcapspace = 2cm]{caption}
\usepackage[scaled=.92]{helvet} % for proper fonts
\usepackage{graphicx} % for EPS use the graphics package instead
\usepackage{balance}  % for useful for balancing the last columns
\usepackage{booktabs} % for pretty table rules
\usepackage{ccicons}  % for Creative Commons citation icons
\usepackage{ragged2e} % for tighter hyphenation

\usepackage{marginnote} 
\usepackage[utf8]{inputenc} % for a UTF8 editor only

\def\plaintitle{SIGCHI Extended Abstracts Sample File: Note Initial
  Caps} 
\def\emptyauthor{}
\def\plainkeywords{Smartphone Overuse Strategies; Mobile App Addiction; App Limits}

\title{App Limits Bar: A Progress of App Limits for Overcoming Smartphone Overuse}

\numberofauthors{1}
\author{%
  \alignauthor{%
    \textbf{Yixian Wu}\\
    \affaddr{School of Artificial Intelligence} \\
    \affaddr{Southeastern University} \\
    \affaddr{Nanjing, 210096, Jiangsu, China}
    \email{213182050@seu.edu.cn} } }

\definecolor{linkColor}{RGB}{6,125,233}
\hypersetup{%
  pdftitle={\plaintitle},
  pdfauthor={\emptyauthor},
  pdfkeywords={\plainkeywords},
  bookmarksnumbered,
  pdfstartview={FitH},
  colorlinks,
  citecolor=black,
  filecolor=black,
  linkcolor=black,
  urlcolor=linkColor,
  breaklinks=true,
}

\begin{document}

%% For the camera ready, use the commands provided by the ACM in the Permission Release Form.
\CopyrightYear{2020}
\setcopyright{rightsretained}
\conferenceinfo{CHI'20,}{April  25--30, 2020, Honolulu, HI, USA}
\isbn{978-1-4503-6819-3/20/04}
\doi{https://doi.org/10.1145/3334480.XXXXXXX}
%% Then override the default copyright message with the \acmcopyright command.
\copyrightinfo{\acmcopyright}

\maketitle

% comment to disable hyphenation (not recommended)
% https://twitter.com/anjirokhan/status/546046683331973120
\RaggedRight{} 

% Do not change the page size or page settings.
\begin{abstract}
Smartphone overuse has many negative effects on human beings. The function App Limits in our phones, which belongs to behavior reinforcement strategies for overcoming problematic smartphone overuse, constrains users by shutting them completely out of an app after a certain period of time. While it has effectiveness to some extent, lacking procedural detection of problematic behavior always brings about anxiety and depression for users. We proposed App Limits Bar (ALB) to progress the traditional App Limits, aiming to mitigate the negative feelings about using App Limits and enhance users' abilities to manage their time on addictive apps. Meanwhile, we took a three-phase user study to answer whether ALB helps users be more content with using the app limits function and whether ALB performs better than traditional App Limits in terms of reducing usage time and open times on addictive apps. Future work will study the best shape of the edge screen for visualization when the front screen faces us.
\end{abstract}

\keywords{\plainkeywords}

% ACM Classfication

\begin{CCSXML}
<ccs2012>
<concept>
<concept_id>10003120.10003121</concept_id>
<concept_desc>Human-centered computing~Human computer interaction (HCI)</concept_desc>
<concept_significance>500</concept_significance>
</concept>
<concept>
<concept_id>10003120.10003121.10003125.10011752</concept_id>
<concept_desc>Human-centered computing~Smartphone overuse</concept_desc>
<concept_significance>300</concept_significance>
</concept>
<concept>
<concept_id>10003120.10003121.10003122.10003334</concept_id>
<concept_desc>Human-centered computing~User studies</concept_desc>
<concept_significance>100</concept_significance>
</concept>
</ccs2012>
\end{CCSXML}

\ccsdesc[500]{Human-centered computing~Human computer interaction (HCI)}
\ccsdesc[300]{Human-centered computing~Smartphone overuse}
\ccsdesc[100]{Human-centered computing~User studies}

% Print the classification codes
% \printccsdesc
% Please use the 2012 Classifiers and see this link to embed them in the text: \url{https://dl.acm.org/ccs/ccs_flat.cfm}

\section{Introduction}
Excessive and problematic use of smartphones, also referred to as smartphone addiction, has potentially harmful effects. Research indicates that musculoskeletal impairment \cite{Alsalameh2019EvaluatingTR}, poor academic performance \cite{Lepp2014TheRB}, anxiety and depression  \cite{Harwood2014ConstantlyC} as well as poor sleep quality  \cite{Chang2014EveningUO} are all negative consequences of smartphone overuse. In recent years, some smartphones allow users to set limited usage time to apps (\emph{e.g.}, Huawei Digital Balance and Iphone App Limits) in order to overcome smartphone addiction. Findings from prior works indicate that smartphone use is reduced as a result of apps regulating use \cite{Busch2020AntecedentsAC}. Nevertheless, notification prompt informing users of remaining time only at the last few minutes may frustrate users if they wasted previous limited time on meaningless contexts of that app. That's because it is obviously late at this moment for them to take action to change their behaviors. Take WeChat as an example, meaninglessly browsing friends’ updates without consciousness occupies a lot during no notification period. Therefore, with the arrival of notification implying that only 5 minutes are left for use, users might feel regretted about their previous improper use of this app and feel anxious about being refused to receive and deal with a more important subsequent message from this app. To solve the mental pressure issue on using the existing app limits function and to change their problematic use of addictive apps, we propose App Limits Bar (ALB) to upgrade the traditional App Limits.

% \marginpar{%
%   \vspace{-45pt} \fbox{%
%     \begin{minipage}{0.925\marginparwidth}
%       \textbf{Good Utilization of the Side Bar} \\
%       \vspace{1pc} \textbf{Preparation:} Do not change the margin
%       dimensions and do not flow the margin text to the
%       next page. \\
%       \vspace{1pc} \textbf{Materials:} The margin box must not intrude
%       or overflow into the header or the footer, or the gutter space
%       between the margin paragraph and the main left column. The text
%       in this text box should remain the same size as the body
%       text. Use the \texttt{{\textbackslash}vspace{}} command to set
%       the margin
%       note's position. \\
%       \vspace{1pc} \textbf{Images \& Figures:} Practically anything
%       can be put in the margin if it fits. Use the
%       \texttt{{\textbackslash}marginparwidth} constant to set the
%       width of the figure, table, minipage, or whatever you are trying
%       to fit in this skinny space.
%     \end{minipage}}\label{sec:sidebar} }

App Limits Bar (\textbf{ALB}) is designed to progress the function App Limits in our mobile phones. ALB can present usage time or remaining time for use during the whole using process, which helps users manage their limited use time better. Meanwhile, we propose to have the left edge of the phone to be a display screen for ALB considering the limited screen of smartphones. In addition, signals representing compulsive open times in the using process are also visualized as Ding \emph{et al.} \cite{Ding2016BeyondSO} proved that compulsive open time is also a good indicator of app addiction.

\section{Related Work}

There are three \cite{Busch2020AntecedentsAC} categories of corrective measures for problematic smartphone overuse: information-enhancing strategies, capacity-enhancing strategies and behavior reinforcement strategies. The proposed ALB belongs to the mixture of the latter two strategies.

Several \textbf{information-enhancing strategies} include direct warnings \cite{Barnes2019MobileUU}, documentaries and public campaigns emphasizing the severity and risk of problematic smartphone use \cite{Chen2017GenderDI} and educational programs and guidelines for the young \cite{Cho2017InfluenceOS}. The limitation of information-enhancing strategies is that it probably won't influence those who have already fully informed about the negative effects of smartphone overuse \cite{Kwon2016ExcessiveDO}. In addition, it is difficult for users to quit their addictive behavior voluntarily only via information-enhancing strategies \cite{Busch2020AntecedentsAC}.

\textbf{Behavior reinforcement strategies} refrain users from accessing smartphone after a certain period of time or by making smartphone use more difficult (\emph{e.g.}, \cite{Cho2017InfluenceOS}). The existing function App Limits is designed based on this logic. While it does restrict smartphone overuse to some extent, App Limits still contains a few imperfections. For instance, lacking procedural detection causes inappropriate management of available usage time, and thus brings about regrets and depression. Inspired by the logic of capacity-enhancing strategies which aims at reinforcing self-disciplinary and rational management abilities  \cite{Kwon2016ExcessiveDO}, tools are supposed to build in features that timely detect the problematic behavior \cite{Shin2013AutomaticallyDP} for users to make adjustments. Our proposed App Limit Bar applies that.

In terms of progress bar design, we apply Harrison \emph{et al.}’s visually augmented progress bar \cite{Harrison2010FasterPB} that makes processes appear 11 \% faster visually, prompting users to feel urgent in the last several minutes for quick preparation of withdrawing themselves from the current app. 

In addition, similar work that Wang \emph{et al.} \cite{Wang2019ThePP} conducted compared point-of-choice prompt with the always-on progress bar on PC for sedentary behavior change, which has a reference influence on our study design.
    
\section{Prototype}

\begin{figure}
  \includegraphics[width=0.9\columnwidth]{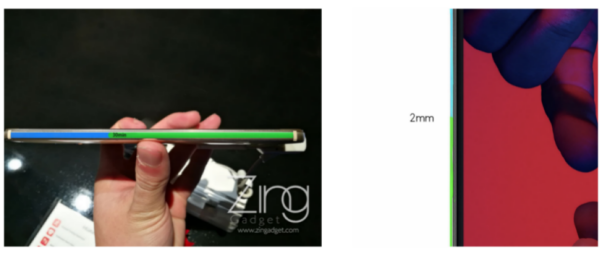}
  \caption{Appearance of App Limits Bar (ALB).}~\label{fig:AppearanceOfALB}
\end{figure}

\begin{marginfigure}[6pc]
  \begin{minipage}{\marginparwidth}
    \centering
    \includegraphics[width=0.7\marginparwidth]{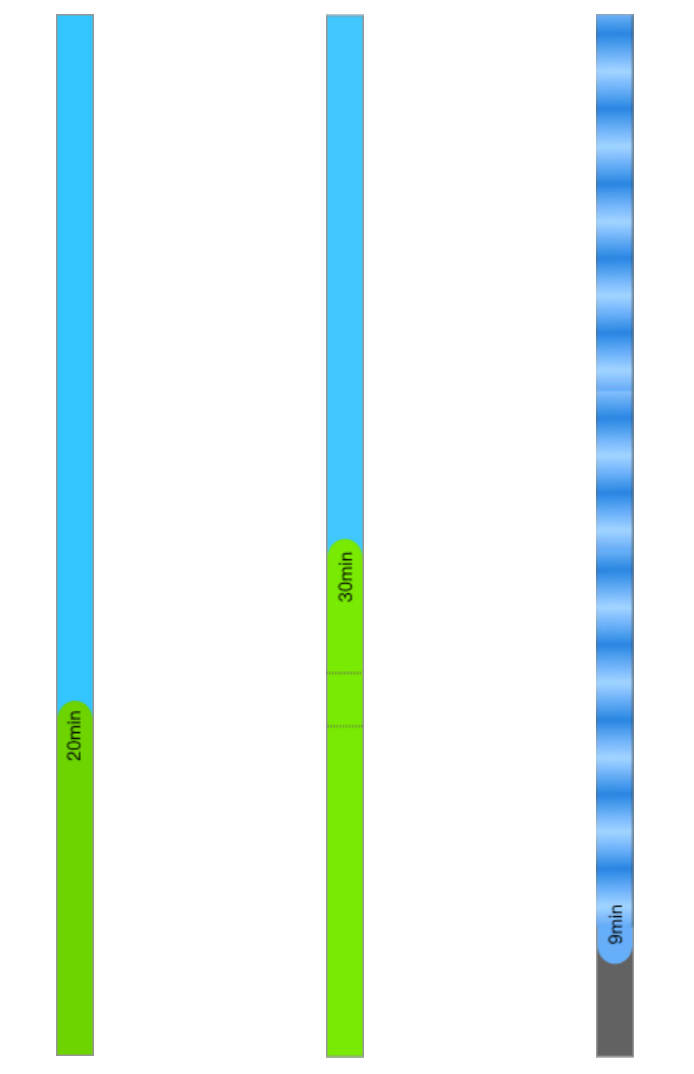}
    \caption{Left: visualization of usage time (green bar) and remaining time (blue bar). Middle: visualization of open times (two imaginary lines). Right: visually augmented bar.}~\label{fig:ProgressBar}
  \end{minipage}
\end{marginfigure}

In order to test the feasibility of our App Limits Bar (ALB), we built a prototype device using the following components:

\begin{itemize}
    \item [1)] An Iphone 11 or other IOS smartphone devices that contain App Limits function.
    \item [2)] A piece of external, curved and multi-touch screen connected with the phone. The display screen is modified to half-terete shape, whose radius is 2 millimeters long, and is matched with the left edge of the phone (see Figure~\ref{fig:AppearanceOfALB}).
\end{itemize}
    
\subsection{Progress Bar}

To achieve ALB function, we are required to develop a progress bar on this external and multi-touch screen, which includes the following functions:

\begin{itemize}
    \item [1)] Record the usage time and open time of the addictive apps. A green bar is designed for recording the usage time while a blue bar is for the left time. The default setting is that the green bar is on the blue one (see Figure~\ref{fig:ProgressBar}). Meanwhile, a designed imaginary line is visualization of the open time (see Figure~\ref{fig:ProgressBar} (middle)).
    \item [2)] A visually augmented bar (see Figure~\ref{fig:ProgressBar} (right)) is to be displayed when the last 10 minutes left.
    \item [3)] A horizontal black bar will be displayed when users slide down the screen versus the bar will be canceled when users slide up (see Figure~\ref{fig:TimeFrom0} and Figure~\ref{fig:TimeFrom0Recovery}).
    \item [4)] To set blue bar is on the green bar, tap twice on the blue bar and vice versa.
\end{itemize}

% \marginpar{%
%   \includegraphics[width=\marginparwidth]{myfile}
%   \captionof{figure}{The caption}
% }

% \begin{table}
%   \centering
%   \begin{tabular}{l r r r}
%     % \toprule
%     & & \multicolumn{2}{c}{\small{\textbf{Test Conditions}}} \\
%     \cmidrule(r){3-4}
%     {\small\textit{Name}}
%     & {\small \textit{First}}
%       & {\small \textit{Second}}
%     & {\small \textit{Final}} \\
%     \midrule
%     Marsden & 223.0 & 44 & 432,321 \\
%     Nass & 22.2 & 16 & 234,333 \\
%     Borriello & 22.9 & 11 & 93,123 \\
%     Karat & 34.9 & 2200 & 103,322 \\
%     % \bottomrule
%   \end{tabular}
%   \caption{Table captions should be placed below the table. We
%     recommend table lines be 1 point, 25\% black. Minimize use of
%     table grid lines.}~\label{tab:table1}
% \end{table}

\section{Study}

This user study is designed to test the effect of ALB. We are intended to answer the following questions by this study:

\begin{itemize}
    \item [1)] Dose ALB make users reduce the usage time and open times cost on addictive apps compared with no limits and traditional App Limits?
    \item [2)] Dose ALB reduce the frequency of choosing "no limits today" compared with the traditional App Limits?
    \item [3)] Do users have more positive feelings about using ALB than traditional App Limits?
\end{itemize}

Question 1 \& 2 are designed to explain whether users manage their limited usage time better than using traditional App Limits. Question 3 is designed to figure out whether ALB reduces negative feelings about using app limits function for users.

\subsection{Participant Selection}
We recruited 12 participants aged 18 to 22 who were college students. Of these participants, half were male and half were female to avoid gender bias. All of them have experienced using App Limits on their phones in order to reduce the time spent on addictive apps. 

\subsection{Tutorials of How to use ALB}
Before using ALB, we were required to assist participants to get familiar with all the ALB functions:

\begin{figure}
  \centering
  \includegraphics[width=0.7\columnwidth]{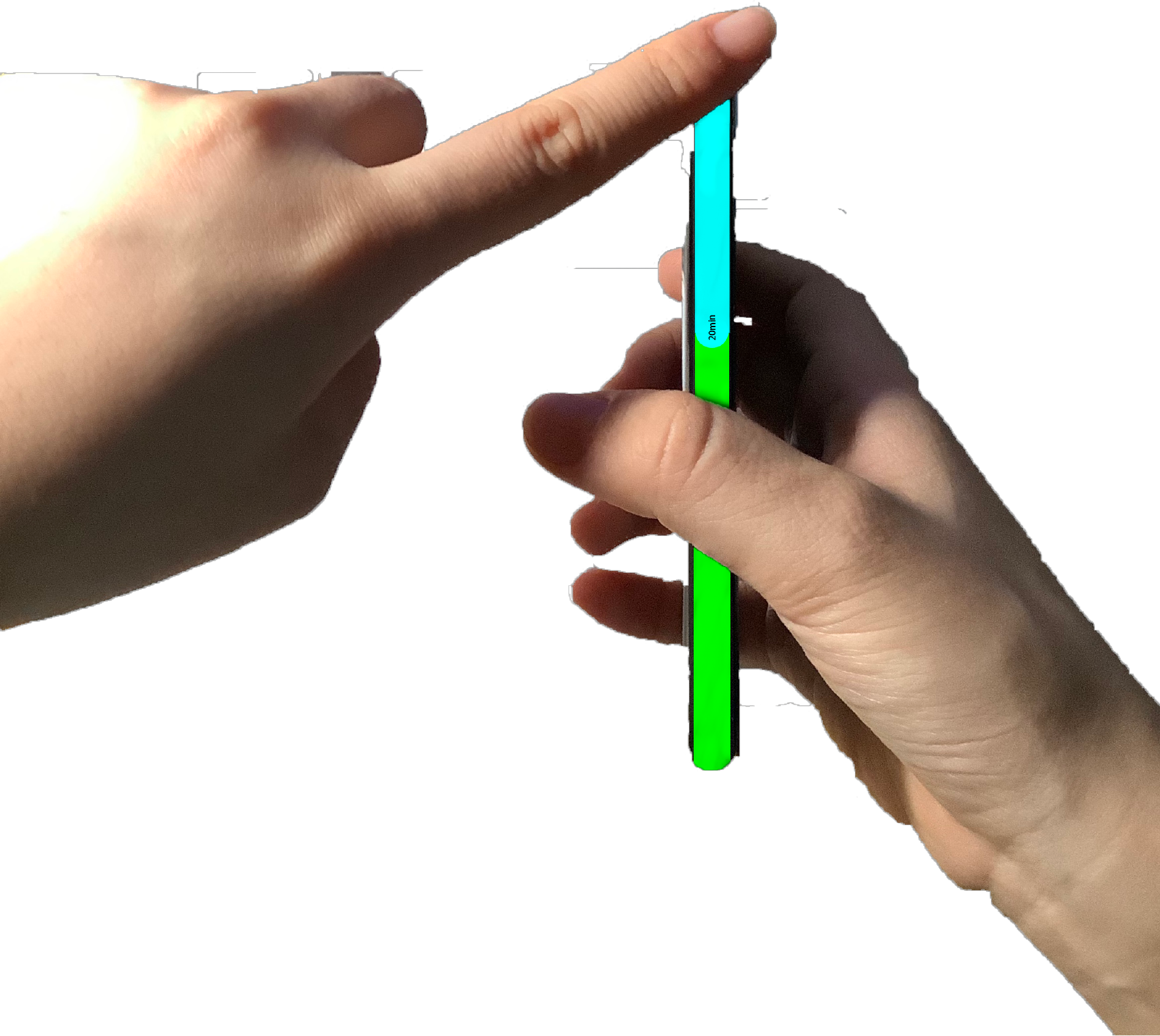}
  \caption{Gesture that evokes touch on the left edge screen.}~\label{fig:HoldPhone}
\end{figure}

\begin{itemize}
    \item Show participants how to set a limited usage time for a certain app using IOS App Limits function.
    
\begin{figure}[h]
  \centering
  \includegraphics[width=0.7\columnwidth]{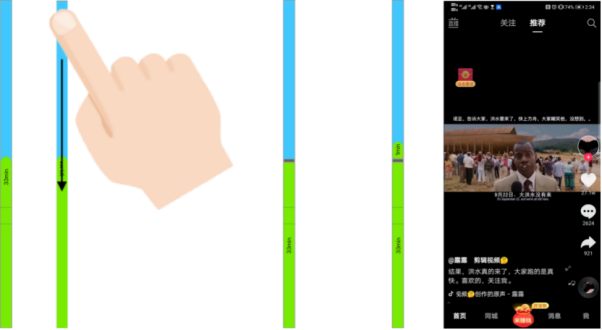}
  \caption{Set a horizontal black bar to time from zero.}~\label{fig:TimeFrom0}
\end{figure}

    \item Open the limited app and browser it for a while. We need to explain that the green bar represents the usage time and the blue bar shows the left time. Meanwhile, participants are supposed to be taught that in order to display a countdown, they need to hold the phone sideways, making the left edge screen upward (see Figure~\ref{fig:HoldPhone}), and then tap the blue bar twice. If they want to recover to show the usage time, hold the phone as mentioned before and tap the green bar twice. In addition, participants should be informed that once they open the app, there will be an imaginary line in the position of the usage time progress bar.
    \item Participants can make the progress bar time from zero instead of continuous accumulation by holding the phone sideways, making the left edge screen upward, and then sliding down on the edge screen with a figure to set a horizontal black bar (see Figure~\ref{fig:TimeFrom0}). In terms of recovery, slide up to delete the latest bar (see Figure~\ref{fig:TimeFrom0Recovery}).

    \item When the blue bar remains last 10 minutes, this 10-minute part will be full of all the edge screen and display final countdown (see Figure~\ref{fig:ProgressBar} (Right)). The pure blue bar will become a visually augmented bar to make users feel urgent. In the end, the point-of-choice prompt will display on the whole screen to inform users that the available usage time is over. Participants can choose either "OK" or "no limits today". If it is "no limits today", participants will be allowed to use this app with no time limits for the rest of the day.
    
\begin{figure}[h]
  \centering
  \includegraphics[width=0.5\columnwidth]{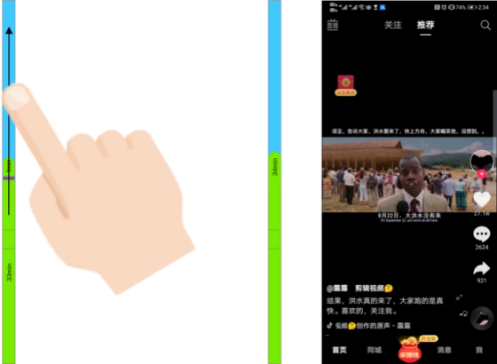}
  \caption{Slide up to delete the latest horizontal black bar.}~\label{fig:TimeFrom0Recovery}
\end{figure}

\end{itemize}

\subsection{Study Design}
In order to answer the research questions above, we ran the study in three phases, which cost three weeks in total:

The first phase is to help participants figure out which apps they are addicted to. We provided an Iphone 11 for each participant and took away their own phones for six days (week 1 from Monday to Saturday). Each participant had downloaded their commonly-used apps on the provided phones ahead of the experiment. During the first phase (week 1 from Monday to Saturday), they used this new phone without any limit. Usage time and open time frequency per day for each app were recorded in the built-in settings of their Iphone. When the first phase was over, each participant was required to review their usage time per day for each app and chose several addictive apps for themselves on the following Sunday.

Finding the specific addictive apps for each participant, phase two and phase three are designed to compare the effect of ALB with the traditional app limit function.

As a preparation for the second phase, participants were asked to use the built-in App Limit function to restrict all the addictive apps they selected in the first phase during the following six days (week 2 from Monday to Saturday). While the participants can set time for each addictive app according to their own will, the restricted time per day for each addictive app should be less than the maximum daily usage time of that app in the first phase. Usage time per day, open time frequency per day, and how many times they choose "no limits today" in these six days were recorded. What needs to be emphasized is that participants cannot be informed that they are going to be recorded the chosen frequency of "no limits today" in the experimental days.

Prior to the last phase, we assisted all the participants to get familiar with ALB (see \emph{Tutorials of How to use ALB}) on Sunday of week 2. In the last phase (week 3 from Monday to Saturday), each participant were asked to use the ALB to restrict all the addictive apps (NOTE. The restricted usage time and apps were the same as the second phase). Usage time per day, open time frequency per day, and how many times they choose "no limits today" in these six days were also recorded. In addition, that we are going to record the chosen frequency of "no limits today" in these six days cannot be known by participants either.

On the last Sunday, each participant was interviewed by us, answering the following questions:

\begin{itemize}
    \item [1)] Do you feel less frustrated and anxious when the limited time is up during phase three compared with phase two? (Yes or No)
    \item [2)] Do you think the App Limits Bar helps you reduce meaningless usage time on addictive apps compared with the traditional App Limits function? (Yes or No)
    \item [3)] Do you think your management and self-regulation abilities on app addiction have improved during phase three compared with phase two? (Yes or No)
\end{itemize}

\subsection{Data Collection and Processing}

For each participant, we drew the line chart of usage time per day of each app distributed in the six days from three phases for the convenience of comparison. Open times per day for each app were made the same statistics as the usage time. The chosen frequencies of "no limits today" from three phases were also compared. In addition, we also did statistics on the answers to each interview question.

% \marginpar{\vspace{-23pc}So long as you don't type outside the right
%   margin or bleed into the gutter, it's okay to put annotations over
%   here on the left, too; this annotation is near Hawaii. You'll have
%   to manually align the margin paragraphs to your \LaTeX\ floats using
%   the \texttt{{\textbackslash}vspace{}} command.}

% \begin{margintable}[1pc]
%   \begin{minipage}{\marginparwidth}
%     \centering
%     \begin{tabular}{r r l}
%       & {\small \textbf{First}}
%       & {\small \textbf{Location}} \\
%       \toprule
%       Child & 22.5 & Melbourne \\
%       Adult & 22.0 & Bogot\'a \\
%       \midrule
%       Gene & 22.0 & Palo Alto \\
%       John & 34.5 & Minneapolis \\
%       \bottomrule
%     \end{tabular}
%     \caption{A simple narrow table in the left margin
%       space.}
%   \label{tab:table2}
%   \end{minipage}
% \end{margintable}

\section{Limitation}
For our design, we mainly focused on ALB's effectiveness. Therefore, edge screen design has not been well explored. Further work will research whether the half-terete edge screen is able to display ALB best when the front screen faces us compared with other possible shape design on the edge screen.

\balance{} 

\bibliographystyle{SIGCHI-Reference-Format}
\bibliography{main}
\input{main.bbl}

\end{document}